\documentstyle[aps,prd,epsfig]{revtex}
\newcommand{\s}{$\sigma_{tot}^{pp}$ }

\tightenlines

\begin{document}

\title{On the total cross section extrapolations to cosmic-ray energies}
\author{E. G. S. Luna and M. J. Menon }
\address{Instituto de F\'{\i}sica ``Gleb Wataghin''\\
Universidade Estadual de Campinas, UNICAMP\\
13083-970, Campinas, SP, Brazil}

\maketitle

\begin{abstract}

 We analyze $pp$ total cross section data from
accelerator to cosmic-ray energies and show that the experimental information
presently available indicate two different scenarios at the highest energies.
One of them accommodates the predictions of the majority of models and the
other one shows agreement with a multiple diffraction model recently developed
and improved. We argue that these results  may bring a kind of warning against
some possible precipitated attempts to constraint cosmic-ray estimations from
analyzes of accelerator data.

\end{abstract}

\vspace{0.3cm}
PACS numbers: 13.85.-t, 13.85.Lg, 13.85.Tp
\vspace{0.3cm}

Some years ago, a successfull description of $pp$ elastic scattering data in
the interval $13.8 \le \sqrt s \le 62.5$ GeV was obtained through a
multiple diffraction model \cite{mm}, hereafter referred as MDM. Extrapolations
of the predicted \s showed agreement with the highest quoted values of the
cosmic-ray experiments. In particular, the model predicts
$\sigma_{tot}^{pp}(\sqrt s = 16\ \textnormal{TeV}) = 147$ mb (without
error estimated) \cite{mm}. To our knowledge this is the only
phenomenological approach that indicates that the cross section may rise
faster with the energy than generally expected from ``standard'' model
predictions \cite{dl,dgp,bsw,block} and some fit results 
\cite{ua4/2}. 

Recently, P\'erez-Peraza {\it et al.} introduced some improvements in the model
predictions by determining confident error bands, through a forecasting
regression analysis \cite{perez}. According to the authors, the extrapolations
present so large error bands that the predictions show agreement with all the
cosmic-ray results, including those that are incompatible among themselves. Moreover, 
the authors show that taking account of the $pp$ data in the interval $13.8
\le  \sqrt s \le 62.5$ GeV together with $\overline{p}p$ data at $546$ GeV
and $1.8$ TeV, the predictions resulting from a new fit of the model free
parameters present a very narrow error band and are compatible with the
smallest quoted values of the cosmic-ray experiments. Since this seems to be
in agreement with ``standard'' models and fit results
\cite{dl,dgp,bsw,block,ua4/2}, the authors conclude that ``extrapolations from
accelerator data should be used to constraint cosmic-ray estimations''
\cite{perez}, which was also suggested in \cite{velasco} and done in practice
in \cite{block}.

However some aspects of the approach \cite{perez} must be commented. Firstly,
reading from the plot in Fig. 2 of this reference we can infer
$\sigma_{tot}^{pp}(\sqrt s = 16\ \textnormal{TeV}) \sim 147 \pm 37$ mb.
Although this is a large error band ($\sim 25 \% $) the results still favor
(at least qualitatively) the highest quoted values of the cosmic-ray
experiments, as shown in the same Figure (this is not what the authors
claimed). Secondly, in the context of a model approach,
simultaneous analysis of $pp$ and $\overline{p}p$ scattering 
should take analyticity and crossing into account, what was not done
in \cite{perez} and also in \cite{velasco}. At last, the assumption that $pp$
and $\overline{p}p$ interactions are the same even at $546$ GeV is a strong
{\it ad hoc} hypothesis and since  $pp$ and $\overline{p}p$ interactions are
different in the ISR energy region, one should expect the use of 
$\sigma_{tot}^{\overline{p}p}$ and not $\sigma_{tot}^{pp}$ in this interval.

In order to attempt to bring some insight on this issue, we performed a simple
model-independent analysis of the same $pp$ accelerator data used in \cite{mm}
and \cite{perez,velasco}, but
including cosmic-ray information presently available. As reviewed in
some detail in \cite{mm}, at cosmic-ray energies ($\sqrt s : 6 - 40$ TeV)
we can identify two distinct set of estimations, one represented by the results
of the Fly's Eye Collaboration \cite{fly} together with those by the Akeno
Collaboration \cite{akeno}, and other by the results of Gaisser, Sukhatme, Yodh
(GSY)\cite{gsy} with those by Nikolaev \cite{niko}. All these data and
estimations are displayed in Fig. 1 (Akeno results were read from the graph
of Fig. 4 in Ref. \cite{akeno}).

The discrepancies between these two sets of cosmic-ray information comes from
both,
the uncertainties in the determination of the production cross
section $\sigma_{p-air}^{prod}$, and the uncertainties in
the connection between $\sigma_{p-air}^{prod}$ and the $pp$ cross section, as
discussed in detail by Engel, Gaisser, Lipari and Stanev \cite{engel}: all the
cosmic-ray estimations are strongly model dependent.
One important point is the fact that the results by Nikolaev and GSY have never
been directly critized in the literature.

For these reasons, we shall investigate the behavior of the
total cross section by taking into account  the discrepancies that
characterize the cosmic ray information. Also, since the base of our
discussion is the predictions of the MDM,  we shall consider the same set of
$pp$ accelerator data used in references \cite{mm} and \cite{perez,velasco}.
In order to accomplish that, we select {\it three ensembles of data and
experimental information} with the following notation:

\begin{itemize}
\item  Ensemble I: $pp$ accelerator data \cite{pp},
\item Ensemble II:  $pp$ accelerator data + Akeno + Fly's Eye 
\cite{pp,fly,akeno},
\item Ensemble III: $pp$ accelerator data + Nikolaev +
GSY \cite{pp,gsy,niko}.
\end{itemize}

\begin{figure}
\begin{center}
\epsfig{figure=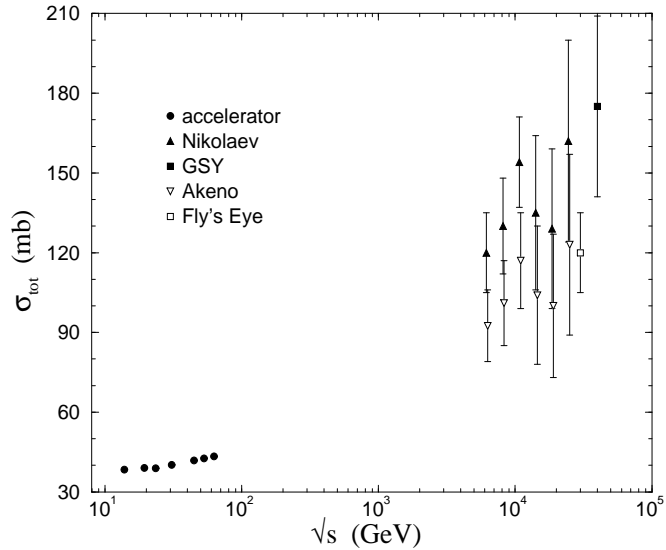,width=10.0cm,height=8.5cm}
\caption{Proton-proton total cross sections: accelerator data
\protect\cite{pp} and cosmic-ray estimations
\protect\cite{fly,akeno,gsy,niko}}
\end{center}
\end{figure}

Different analytical parametrizations may fit all these
ensembles with quite good statistical results, as shown, for example, in
\cite{alm}. In this communication we shall choose the simplest form, which
corresponds to the same analytical function used to fit the free parameters,
depending on the energy, in the MDM. As explained in \cite{mm} this concerns a
polynomial of second degree in $\ln s$. In terms of \s this choice satisfies
the Froissart-Martin bound,

\begin{equation}
\sigma_{tot}^{pp}= A+B\ln \frac{s}{s_0} +C\left[ \ln \frac{s}{s_0}\right]^{2},
\end{equation}
where $A$, $B$, $C$ are free parameters and $s_0 = 1$ GeV$^2$.

It should be stressed that in \cite{mm} the fits involved only the
differential cross section and the $\rho$ parameter data (input information).
The total cross section was {\it predicted} from parametrizations of
essentially two free parameters depending on the energy, one associated with
the hadronic form factor and the other with the average elementary scattering
amplitude. Since these parameters enter in the eikonal (in the momentum
transfer space), the final result for \s should not be expected to be the same
as that obtained with Eq. (1). In addition, it should be noted that other
parametrizations including terms like $ (\ln s)^{D}$  and $Rs^{-1/2}$ ($D$ and
$R$ free parameters) lead to a faster rise of $\sigma_{tot}^{pp}(s)$ than that
obtained through Eq. (1), as shown in \cite{alm}. In this sense the above
choice clearly constraint the rise of  $\sigma_{tot}^{pp}$ and shall be
considered as a lower limit (allowing some connection with the
parametrizations of the MDM).

The fits have been performed using the CERN-MINUIT routine \cite{minuit}. The
corresponding values of the free parameters and statistical information
about each fit are displayed in Table 1. Errors from the free parameters were
propagated to $\sigma_{tot}^{pp}(s)$ through the usual formulas, taking
account of both variances and covariances \cite{bevington}.
The results with the corresponding error bands ($\pm$ one standard
deviation) for ensembles I, II and III are shown in Figs. 2, 3,  and 4,
respectively.

\begin{minipage}{16.0cm}
\begin{table}
\begin{center}
Table 1. Values of the parameters $A$, $B$ and $C$ in Equation $(1)$
(all in mb) and the $\chi^2$ per degree of freedom in each fit to Ensembles
I, II and III. 
\begin{tabular}{cccc}
\hline
Ensemble:       &     I     &   II        &    III      \\
\hline
$A$             &  47.32    &     45.78   &     49.27    \\
$B$             &  -3.809   &   -3.315    &      -4.435  \\
$C$             &   0.4042  &    0.3654   &      0.4534  \\
$\chi^2$        &   8.654   &    11.75    &      15.04    \\
$d.o.f.$        &   4       &     11      &     11        \\
$\chi^2/d.o.f.$ &   2.16    &    1.07     &     1.37     \\
\hline
\end{tabular}
\end{center}
\end{table}
\end{minipage}

From Fig. 2, the results with the ensemble I lie between the two sets
of discrepant cosmic-ray estimations. It corresponds to extrapolations
from a ``pure'' accelerator information in the interval investigated.

On the other hand, if one or the other cosmic-ray sets are included the
situation changes, as shown in Figs. 3 and 4. Taking into account the error
bands (one standard deviation), the two fits become discrepant 
above $\sim 1$ TeV and they are certainly incompatible above $5$ TeV.
We understand this result as a {\it statistical evidence for the possibility
of two distinct scenarios above} $\sqrt s \sim 5$ TeV.

In particular, from each ensemble, the interpolated results at $\sqrt s = 14$
TeV are $\sigma_{tot}^{pp} = 125 \pm 7$ mb (I),
$\sigma_{tot}^{pp} = 119 \pm 5$ mb (II), and
$\sigma_{tot}^{pp} = 133 \pm 6$ mb (III).
In Ref. \cite{alm} four different parametrizations were used
to fit Ensembles II and III and the two scenarios are also inferred.
For example, taking into account the 4 parametrizations (without error
estimations) we can infer from Ensemble II the value $\sigma_{tot}^{pp}(14\
\textnormal{TeV}) = 113 \pm 5$ mb and from Ensemble III,
$\sigma_{tot}^{pp}(14\ \textnormal{TeV}) = 140 \pm 7$ mb. At lower values of
energy, namely $\sqrt s$ in the region $10 - 100$ GeV, there is no
significant distinction between the four parametrizations.

\vspace{-0.9cm}
\begin{figure}
\begin{center}
\epsfig{figure=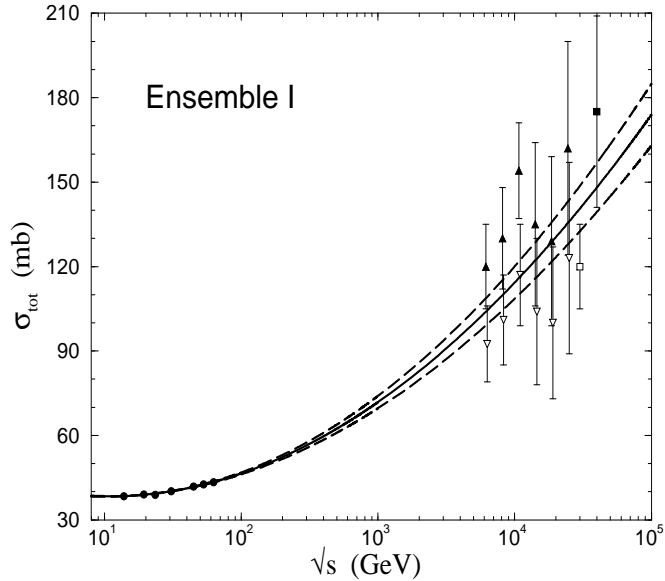,width=10.0cm,height=9.0cm}
\vspace{-0.6cm}
\caption{Parametrization to Ensemble I: \protect$pp$ accelerator data
\protect\cite{pp}.}
\end{center}
\end{figure}
\vspace{-0.4cm}

\vspace{-0.9cm}
\begin{figure}
\begin{center}
\epsfig{figure=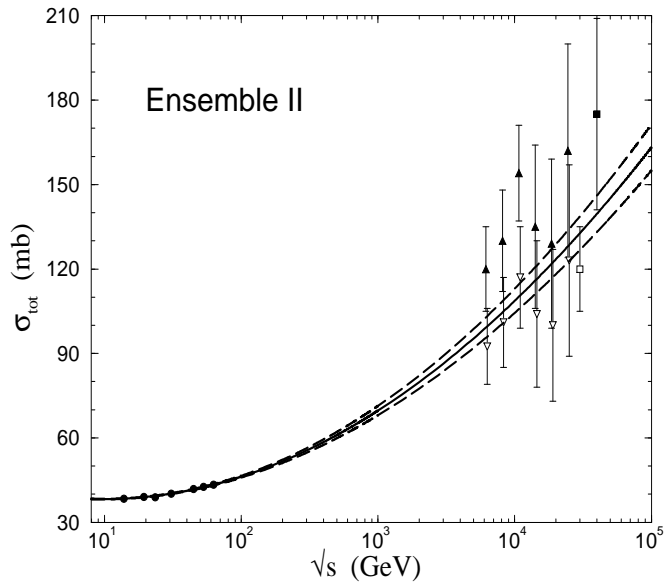,width=10.0cm,height=9.0cm}
\vspace{-0.6cm}
\caption{Parametrization to Ensemble II: \protect$pp$ accelerator data + Akeno +
Fly's Eye \protect\cite{pp,fly,akeno}. }
\end{center}
\end{figure}
\vspace{-0.4cm}

\vspace{-0.9cm}
\begin{figure}
\begin{center}
\epsfig{figure=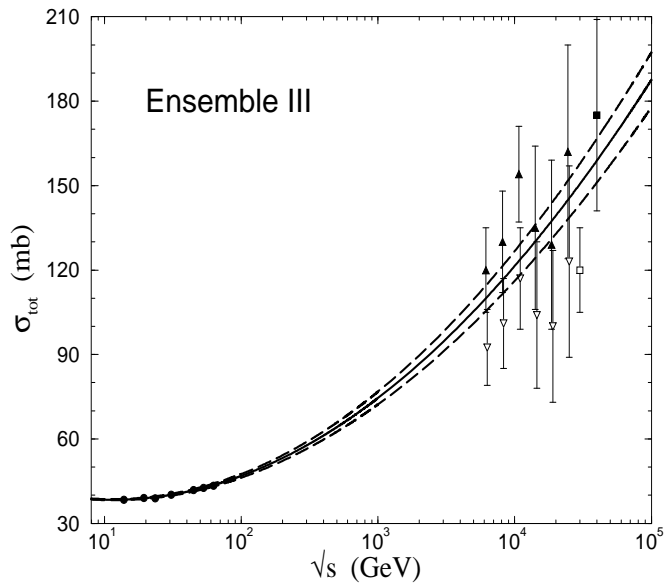,width=10.0cm,height=9.0cm}
\vspace{-0.6cm}
\caption{Parametrization to Ensemble III: \protect$pp$ accelerator data + Nikolaev +
GSY \protect\cite{pp,gsy,niko}.}
\end{center}
\end{figure}

Ensemble II clearly accommodates the predictions from the majority of models
and in general is considered as ``the cosmic-ray results''. On the other hand,
if the estimations in Ensemble III are the correct ones,
the experimental information on \s presently available indicate that the
$pp$ cross section rises faster with the energy than generally expected or
assumed and, in this case, the rise is well described by the MDM.
At this point it should be stressed that, due to its phenomenological basis,
this model can not provide a ``microscopic'' explanation for this phenomena
(only, perhaps, some naive geometrical arguments) and this puts
limitations in the understanding of a ``fast rising''.

However, it should be remembered that several cosmic-ray experiments point out
to the possibility of new phenomena in $pp$ collisions at $\sqrt s \sim 5$ TeV
\cite{chacaltayapamir} and this has never been investigated by accelerators.
Moreover, recently, more confident analyzes stress the particular
features of Centauro events, when compared with normal events
observed in emulsion chamber experiments \cite{shibuya}. Since the bulk of the
cosmic-ray showers concerns $pp$ and not $\overline{p}p$ interactions, these
``exotic'' events may represent an additional contribution to the $pp$
channel, resulting in a faster increase of the \s, as claimed
in \cite{conjecture}.

Certainly, these puzzles shall be resolved by the CERN Large Hadron
Collider (LHC) and even the BNL  Relativistic Heavy Ion Collider (RHIC).
But before this, we understand that the results presented here may serve as  a
kind of warning against possible precipitated attempts to constraint cosmic-ray
estimations from analyzes of accelerator data.

Since in this communication our central goal was to discuss the predictions of
the MDM, the same set of $pp$ accelerator data was used. An extend analysis
with all experimental data and estimations available, including $\overline{p}p$
scattering, is in course. \vspace{0.5cm}

We are grateful to Fapesp for financial support and to R.F. \'Avila for
discussions. M.J.M. is also thankful to N.N. Nikolaev for discussions
and CNPq for financial support.


\begin{thebibliography}{9}

\bibitem{mm} A. F. Martini and M. J. Menon, Phys. Rev. D {\bf 56},
4338 (1997).


\bibitem{dl} A. Donnachie and P. V. Landshoff, Z. Phys. C {\bf 2}, 55 (1979);
Phys. Lett. B {\bf 387}, 637 (1996).

\bibitem{dgp} P. Desgrolard, M. Giffon, and E. Predazzi, Z. Phys. C {\bf 63},
241 (1994).

\bibitem{bsw} C. Bourrely, J. Soffer, and T. T. Wu, Nucl. Phys. {\bf B247}, 15
(1984); Z. Phys. C {\bf 37}, 369 (1988).

\bibitem{block} M. M. Block, F. Halzen, and T. Stanev, Phys. Rev. D {\bf 62},
077501 (200); Phys. Rev. Lett. {\bf 83}, 4926 (1999).

\bibitem{ua4/2}  C. Augier {\it et al.} (UA4/2 Collaboration), Phys. Lett. B
{\bf 315}, 503 (1993).


\bibitem{perez} J. P\'erez-Peraza {\it et al.}, e-Print Archive:
hep-ph/0011167 (2000).

\bibitem{velasco} J. Velasco
{\it et al}, in {\it 26th International Cosmic Ray Conference}, 1999, Vol. 1,
p. 198; e-Print Archive: hep-ph/9910484.


\bibitem{fly} Fly's Eye Collaboration (R. M. Baltrusaitis {\it et al.}), Phys
Rev. Lett. {\bf52}, 1380 (1984).

\bibitem{akeno} Akeno Collaboration (H. Honda {\it et al.}, Phys. Rev. Lett. 
{\bf 70}, 525 (1993).


\bibitem{gsy} T.K. Gaisser, U.P. Sukhatme, and G.B. Yodh, Phys. Rev. D {\bf
36}, 1350 (1987).


\bibitem{niko} N.N. Nikolaev, Phys. Rev. D {\bf 48}, R1904 (1993).


\bibitem{engel} R. Engel, T. K. Gaisser, P. Lipari, and T. Stanev, Phys. Rev.
D {\bf 58}, 014019 (1998).


\bibitem{pp} Accelerator data on \s (the numbers between the squared brackets 
indicate the center-of-mass energy in GeV): A.S. Carrol {\it et al.}, Phys.
Lett. {\bf B 61}, 303 (1976) [13.8]; A.S. Carrol {\it et al.}, Phys. Lett. {\bf
B 80}, 423 (1979) [19.4]; U. Amaldi and K. R. Schubert, Nucl. Phys. {\bf B
166}, 301 (1980) [23.5, 30.7, 44.7, 52.8, 62.5].

\bibitem{alm} R. F. \'Avila, E. G. S. Luna, and M. J. Menon, hep-ph/0105065,
submitted to Braz. J. Phys.

\bibitem{minuit} F. James, {\it MINUIT - Function Minimization and Error
Analysis} - Reference Manual, Version 94.1, CERN - D506 (1994).

\bibitem{bevington} P. R. Bevington and D. K. Robinson, {\it data reduction
and error analysis for the physical sciences}, (Mcgraw-Hill, New York, 1992).

\bibitem{chacaltayapamir} Chacaltaya and Pamir Collaboration, Nucl. Phys. 
{\bf B370}, 365 (1992).

\bibitem{shibuya} S. L. C. Barroso et al., Nucl. Phys. B (Proc. Supp.)
{\bf 75A}, 150 (1999); S. L. C. Barroso, Doctoral Thesis, UNICAMP, Brazil,
2000.

\bibitem{conjecture} M. J. Menon, in {\it International Workshop On Hadron
Physics 98}, edited by E. Ferreira, F. F. S. Cruz and S. S. Avancini (World
Scientific, Singapore, 1999) p. 333; e-Print Archive: hep-ph/9810508.


\end{thebibliography}
\end{document}